\documentclass{eas}
\usepackage{graphicx}
\TitreGlobal{A Universe of Dwarf Galaxies: Observations, Theories, Simulations, 14-18 June, Lyon, France}
\begin{document}

\newcommand{\1}{{~\sc i}}
\newcommand{\2}{{~\sc ii}}
\newcommand{\3}{{~\sc iii}}
\newcommand{\4}{{~\sc iv}}
\newcommand{\5}{{~\sc v}}
\newcommand{\6}{{~\sc vi}}
\newcommand{\kms}{{\,km\,s$^{-1}$}}
\newcommand{\mic}{{\,$\mu$m}}
\newcommand{\sol}{$_\odot$}

\newcommand{\apj}{ApJ}
\newcommand{\aj}{AJ}
\newcommand{\apjl}{ApJL}
\newcommand{\apjs}{ApJS}
\newcommand{\aap}{A\&A}
\newcommand{\aapr}{A\&AR}
\newcommand{\aaps}{A\&AS}
\newcommand{\mnras}{MNRAS}
\newcommand{\araa}{ARAA}
\title{ISM enrichment and local pollution in dwarf galaxies}
\author{D.\ Kunth}\address{Institut d'Astrophysique, Paris, 98 bis Boulevard Arago, F-75014 Paris, France}
\author{V. Lebouteiller}\address{Service d'Astrophysique, L'Orme des Merisiers, CEA/Saclay, 91191 Gif-sur-Yvette, France}
\begin{abstract}
The fate of metals after they are released in starburst episodes is still unclear. What phases of the interstellar medium are involved, in which timescales? Evidence has grown over the past few years that the neutral phase of blue compact dwarf (BCD) galaxies may be metal-deficient as compared to the ionized gas of their H\2\ regions. These results have strong implications for our understanding of the chemical evolution of galaxies. We review here the main results and the main caveats in the abundance determination from far-UV absorption-lines. We also discuss possible scenarios concerning the journey of metals into the interstellar medium, or even their ejection from the galaxy into the intergalactic medium.
\end{abstract}
\runningtitle{ISM enrichment and local pollution in dwarf galaxies}

\maketitle

\section{Introduction}
\label{intro}

Within the chemical downsizing scenario, in which massive galaxies are the first to form stars at a high rate (Cen \& Ostriker 1999), dwarf galaxies can remain chemically unevolved since their formation. It is thus possible that some dwarf galaxies in the nearby Universe still contain pristine gas (Kunth \& Sargent 1986; Kunth {\em et al.\ }1994). 

The family of Blue Compact Dwarf (BCD) galaxies represents a major testbed as it includes the most-metal poor star-forming galaxies known (Kunth \& Ostlin 2000). Most of these galaxies contain young (3-10\,Myr) super stellar clusters that dominate the interstellar medium (ISM) enrichment in $\alpha$-elements. Old stellar populations have been detected in BCDs (e.g., Doublier {\em et al.\ }2001; Aloisi {\em et al.\ }1999; Aloisi {\em et al.\ }2001; Ostlin 2000; Vanzi {\em et al.\ }2000; Hopp {\em et al.\ }2001), suggesting that although "chemically young", these objects are not genuinely young. Nevertheless, only 22\% of the total stellar mass in I\,Zw\,18 could be contributed by older stars which in itself makes this object peculiar per se (Hunt {\em et al.\ }2003).

Although evidence of local enrichment has already been observed (Walsh \& Roy 1993), it is usually associated with the presence of Wolf-Rayet (WR) stars (Walsh \& Roy 1989; Thuan {\em et al.\ }1996; Lopez-Sanchez {\em et al.\ }2007). Brinchmann {\em et al.\ }(2008) confirmed this assumption by analyzing 570 WR galaxies showing larger N/O abundance ratios as compared to normal galaxies. In order to probe the ISM enrichment by starburst episodes, dispersion and mixing timescales have to be better understood. For example, a typical 10\,Myr burst involving $2.5\times10^6$\,M\sol\ of stars at a metallicity of 0.2\,Z\sol\ yields about $10^4$\,M\sol\ into the ISM (using yields of Meynet \& Maeder 2002). Considering a mass of ionized gas of $2.5\times10^5$\,M\sol, an instantaneous local enrichment of the newly produced metals would lead to an unrealistic metallicity of 10\,Z\sol! It is clear that observing the ionized gas only (through the optical-infrared emission-lines) does not provide the complete picture of the journey of heavy elements into the ISM. The neutral envelope notably plays a fundamental role.

The role of the neutral gas upon the chemical evolution of galaxies is twofold. First, it acts a massive reservoir for the molecular hydrogen which fuels star-formation. Accretion episodes provide theoretically enough gas to explain the star-formation rate observed in certain galaxies (e.g., Fraternali \& Binney 2008). The importance of accretion is likely enhanced in dwarf galaxies characterized by a lack of rotation and inefficient cooling flows. Furthermore, as star-formation occurs within giant H\2\ regions, newly produced metals can be driven to high galactic latitudes into the H\1\ halo through the so-called galactic fountains. The dispersion spatial and time scales as well as the mixing time scale are however relatively unknown in dwarf galaxies.

The sample of BCDs for which the comparison was made between chemical abundances in the H\1\ envelope and in the ionized gas of their H\2\ regions is presented in Table\,\ref{tab:sample}. It includes the 2 lowest-metallicity star-forming galaxies known I\,Zw\,18 (Aloisi {\em et al.\ }2003; Lecavelier {\em et al.\ }2004) and SBS\,0335-052 (Thuan {\em et al.\ }2005), as well as more metal-rich objects, NGC\,1705 (Heckman {\em et al.\ }2001), Mrk\,59 (Thuan {\em et al.\ }2002), I\,Zw\,36 (Lebouteiller {\em et al.\ }2004), NGC\,625 (Cannon {\em et al.\ }2004), and Pox\,36 (Lebouteiller {\em et al.\ }2009). In addition, the giant H\2\ region NGC\,604 in M\,33 was investigated by Lebouteiller {\em et al.\ }(2006).

We first explain the strategy for determining abundances in the neutral gas of galaxies (Sect.\,\ref{sec:determination}). One major challenges concerns the absorption-line contamination by stellar photospheres for the bursts older than 10\,Myr (Sect.\,\ref{sec:hicont}). We then review the results in Sect.\,\ref{sec:results} and investigate the effect of depletion on dust grains in Sect.\,\ref{sec:depletion}. Finally, we gather some considerations on the enrichment of the neutral gas in Sect.\,\ref{sec:discussion}.

\begin{table}
\caption{Sample of star-forming objects whose neutral gas was investigated with FUSE. GHIIR stands for giant H\2\ region. [O/H] is defined as $\log({\rm O}/{\rm H})-\log({\rm O}/{\rm H})_\odot$. We use the solar reference of Asplund {\em et al.\ }(2005).}
\label{tab:sample}       
\begin{tabular}{lll}
\hline\noalign{\smallskip}
Object & Type & [O/H] (ionized gas)  \\
\noalign{\smallskip}\hline\noalign{\smallskip}
I\,Zw\,18 & BCD & $-1.45$ \\
SBS\,0335-052 & BCD  & $-1.36$ \\
I\,Zw\,36 & BCD  & $-0.89$ \\
Mark\,59 & BCD & $-0.67$ \\
Pox\,36 & BCD  & $-0.61$ \\
NGC\,625 & BCD  & $-0.47$ \\
NGC\,1705 & BCD   & $-0.45$ \\
\hline
NGC\,604 & GHIIR  & $0.00$\\
\noalign{\smallskip}\hline
\end{tabular}
\end{table}

\section{Determining abundances in the neutral gas}
\label{sec:determination}

Metals from the neutral gas can be observed through resonant lines in the far-ultraviolet (FUV). The most massive stars present in BCDs provide a strong FUV continuum, and are used as lighthouses to probe the foreground interstellar gas. The UV photons cross respectively the stellar atmospheres (absorption of H\1\ and highly ionized species such as C\3, C\4, N\5, ...), the photoionized gas (C\2, C\3, S\3, ...), the neutral ISM (H\1, N\1, O\1, Si\2, P\2, Ar\1, Fe\2, ...), and the Milky Way ISM. Provided the radial velocity of the object is large enough to distinguish it from the Milky Way component, the neutral gas is easily probed in BCDs, which display large amounts of H\1\ gas (Thuan \& Martin 1981). 

\begin{figure}
\rotatebox{-90}{\resizebox{0.75\columnwidth}{!}{  \includegraphics{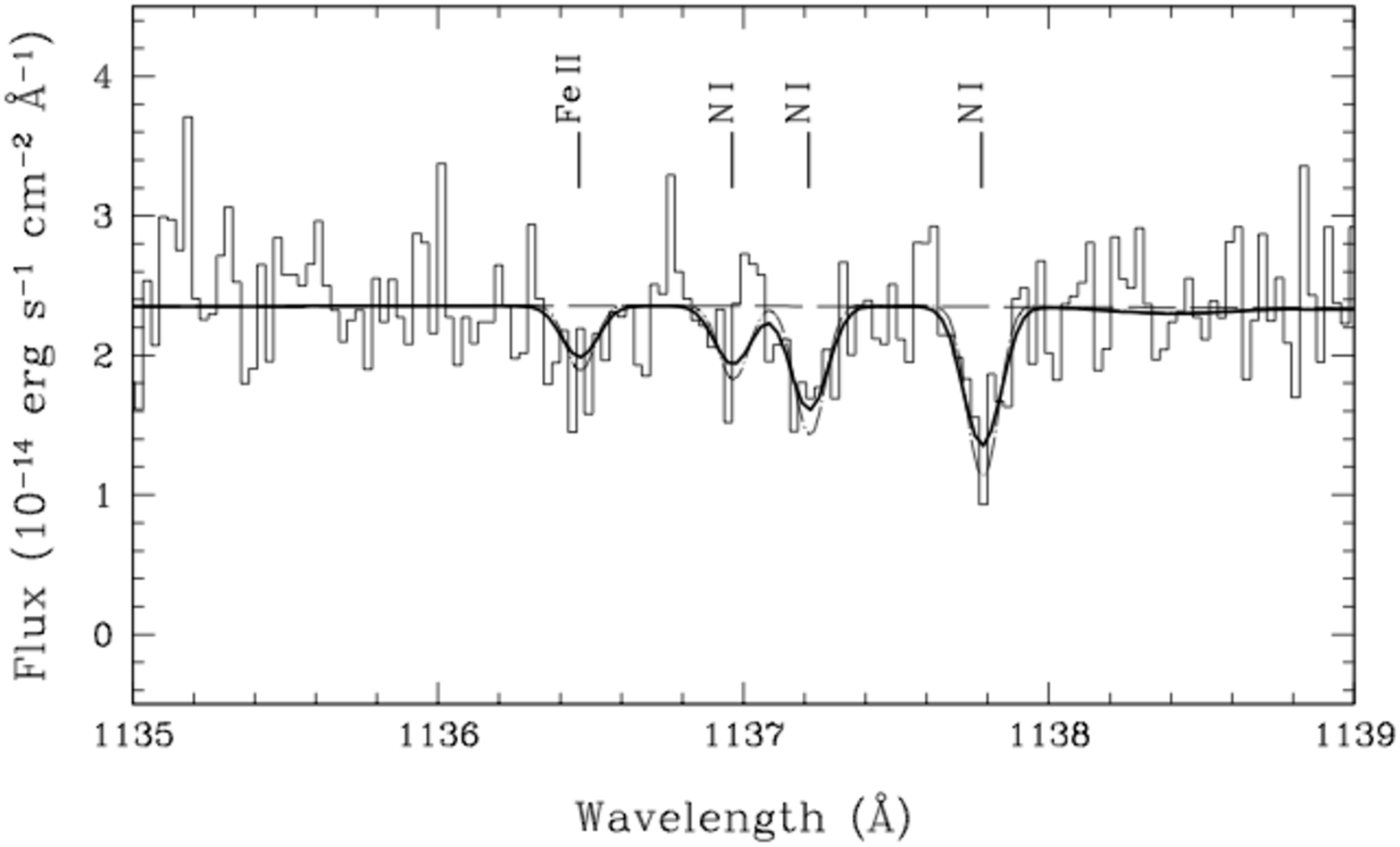} } }
\caption{N\1\ triplet at 1134\,\AA\ toward I\,Zw\,18 (figure taken from Lecavelier {\em et al.\ }2004).}
\label{fig:NItriplet}       
\end{figure}

The FUSE telescope (Moos {\em et al.\ }2000) gave access to the range 900-1200\AA\ at a spectral resolution of 15\,000, allowing for the first time precise determination of metal abundances in the neutral gas, following the pioneering work of Kunth {\em et al.\ }(1994) with the HST/GHRS. Although several neutral gas tracers are available, their use to derive chemical abundances can be challenging for many reasons:
\begin{itemize}
\item The H\1\ lines show prominent damping wings from which the H\1\ column density is easily determined. All the lines from the Lyman series fall within the FUSE range except Ly$\alpha$. The intrinsically large line width makes it often difficult to separate from the Milky Way H\1\ lines. Moreover, stellar atmospheres can contaminate significantly the absorption profile (more details in Sect.\,\ref{sec:hicont}).
\item O\1\ is an important tracer of the $\alpha$-elements produced in massive stars. The ionization potential of O\1\ is very close to that of H\1\ (Table\,\ref{tab:ips}) and the charge exchange between O\2\ and H\1\ is very efficient so that the coupling of the oxygen and hydrogen ionization fractions is quite strong. Yet, practically, O\1\ lines are often saturated due to strong oscillator strengths, making the column density measurement uncertain.
\item N is produced by intermediate-mass to massive stars. The presence of N\1\ in the neutral regions is well established (Table\,\ref{tab:ips}), it can be shown that the applied correction to account for the possible presence of ionized N in the neutral gas is only 0.05\,dex (Sofia \& Jenkins 1998). Finally, little depletion on dust is expected. An example of N\1\ line detection is shown in Fig.\,\ref{fig:NItriplet}.
\item Si\2\ is the main ionization stage of Si in the neutral gas due to the very low ionization potential of Si\1\ (Table\,\ref{tab:ips}). Still, ionization correction are necessary due to the presence of Si\2\ in the ionized gas. Silicon is depleted on grains, though usually less than iron (e.g., Sofia {\em et al.\ }1994; Savage \& Sembach 1996).
\item P\2\ and O\1\ column densities are seen to follow each other in various environments (Lebouteiller {\em et al.\ }2005; but see Jenkins 2009). However, the ionization potential of P\2\ suggests that ionization corrections might be needed due to the presence of P\2\ in the ionized gas.
\item Ar is produced in massive stars and should not be depleted in a low-density H\1\ cloud (Jenkins {\em et al.\ }2000). Ionization correction from low-density, partly-ionized regions of the ISM can be required because of the relatively large photoionization cross section of Ar\1\ (Sofia \& Jenkins 1998). 
\item Fe\2\ is the main ionization stage in the neutral gas. Fe is significantly depleted on dust grains, but a variety of Fe\2\ lines with different oscillator strengths is available within the FUSE range, making the observation of weak (not saturated) lines possible.
\end{itemize}

\begin{table}
\caption{Ionization potentials (H\1 = 13.60\,eV). The species observed with FUSE are in bold. }
\label{tab:ips}       
\begin{tabular}{lll}
\hline\noalign{\smallskip}
    \1   & \2 &  \3 \\
\noalign{\smallskip}\hline\noalign{\smallskip}
\textbf{O\1 = 13.62\,eV} & O\2 = 35.12\,eV & ... \\ 
\textbf{N\1 = 14.53\,eV} & N\2 = 29.60\,eV & ... \\ 
Si\1 = 8.15\,eV & \textbf{Si\2 = 16.34\,eV} & Si\3 = 34.49\,eV \\ 
P\1 = 10.49\,eV & \textbf{P\2 = 19.72\,eV} & P\3 = 30.18\,eV \\ 
\textbf{Ar\1 = 15.76\,eV} & Ar\2 = 27.63\,eV & ... \\ 
Fe\1 = 7.87\,eV & \textbf{Fe\2 = 16.18\,eV} & Fe\3 = 30.65\,eV \\ 
\noalign{\smallskip}\hline
\end{tabular}
\end{table}

Chemical abundances in the neutral gas are simply calculated using the column density ratio of metals to H\1, with some ionization corrections when needed (concerning mostly Ar\1, Si\2, P\2). We refer to the papers on the individual objects for more details.

\section{H\1\ contamination from the stellar atmosphere}
\label{sec:hicont}

The FUV spectrum of most of the BCDs and giant H\2\ regions studied so far with FUSE is dominated by O stars. This is indicative of a starburst age younger than $<10$\,Myr (Robert {\em et al.\ }2003). Because of the high temperature, hydrogen is mostly ionized, and only weak H\1\ photospheric lines can be observed which contribute to the $-$ already saturated $-$ core of the interstellar H\1\ absorption line. Hence FUV spectra toward young starbursts allow a precise determination of the interstellar H\1\ column density, $N$(H\1), by using the damping wings. It must be noted however, that extremely young starburst episodes ($\approx3$\,Myr) show a prominent O\6\ P-Cygni profile which complicates the profile fitting of the Ly$\beta$ line (Lebouteiller {\em et al.\ }2006). 

Old starbursts ($>10$\,Myr) are dominated by B stars. Such stars have overall a more complex FUV spectrum than O stars because the ionization degree is lower. For this reason, photospheric H\1\ lines becomes prominent (Valls-Gabaud 1993). As an illustration, the H\1\ lines from the Lyman series in Mrk\,59 and Pox\,36 show "V-shaped" wings typical from B stars photospheres. While Thuan {\em et al.\ }(2002) circumvented the problem by considering an artificial continuum to fit the profile of the H\1\ lines in Mark\,59, it is also possible to model the stellar absorption.

\begin{figure}
\resizebox{0.99\columnwidth}{!}{  \includegraphics{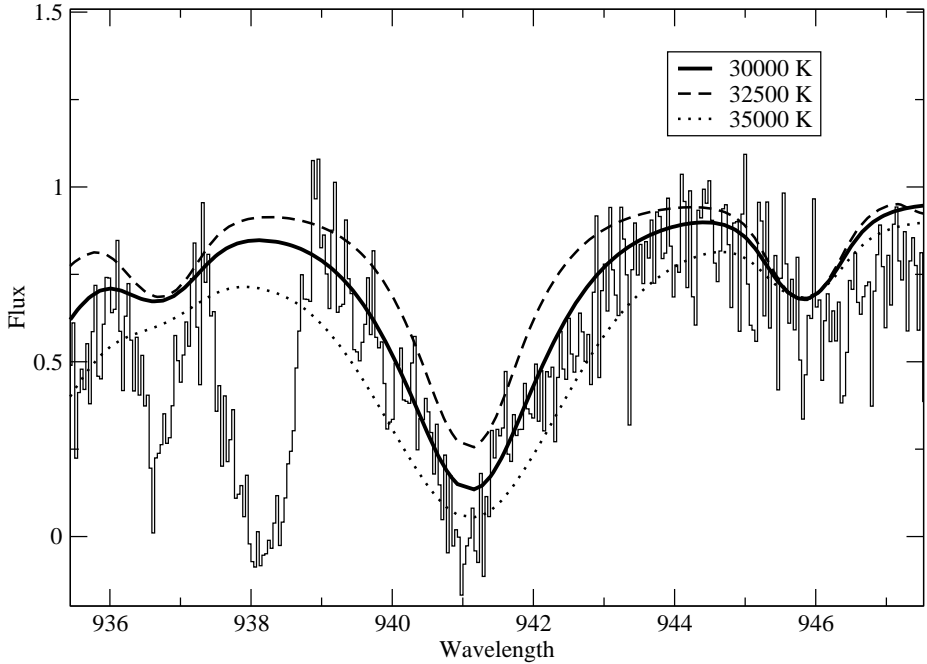} }
\caption{H\1\ Ly$\epsilon$ line toward Pox\,36 (from Lebouteiller {\em et al.\ }2009). The "V-shaped" wings are typical of stellar photosphere absorption and their shape is strongly dependent on the star temperature. The core of the line is saturated and originates from the narrow interstellar H\1. In the case of Pox\,36, the dominant stellar population has $T\approx30\,000$\,K (B0).}
\label{fig:epsilon}       
\end{figure}

In their study of Pox\,36, Lebouteiller {\em et al.\ }(2009) used the TLUSTY models for O and B stars (Lanz \& Hubeny 2003; Lanz \& Hubeny 2007) to reproduce the synthetic spectral continuum. The stellar spectrum was modeled with a single stellar population. It appears that the wings of the stellar H\1\ lines are strongly dependent on the star temperature. The best constraint is given by the high-order line of the Lyman series since the interstellar contribution is saturated with the absence of damping wings (Fig.\,\ref{fig:epsilon}). 

\begin{figure*}
\resizebox{1\columnwidth}{!}{  \includegraphics{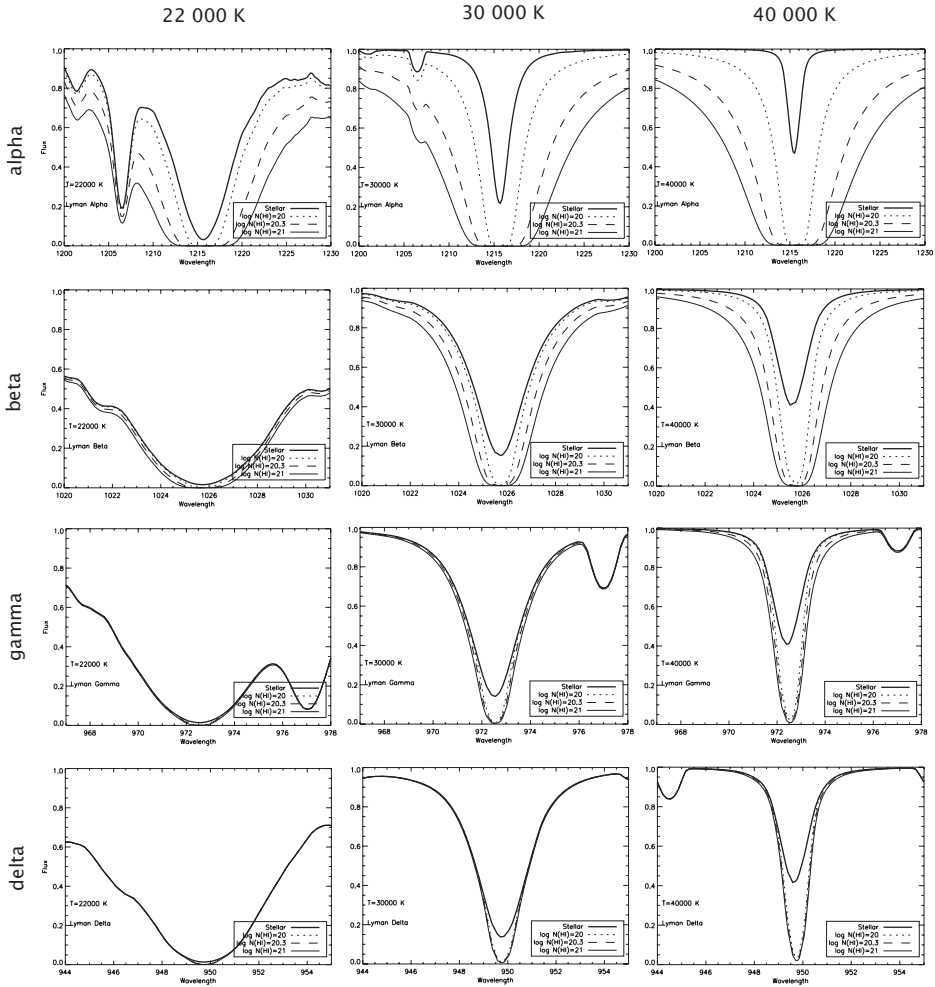} }
\caption{Complicated figure showing the H\1\ Lyman line profiles (stellar and stellar+interstellar) as a function of the star temperature (horizontal), of the Lyman order (vertical), and of the interstellar H\1\ column density (line styles) (figure taken from Lebouteiller {\em et al.\ }2009). For hot stars, the profile of Ly$\alpha$ and Ly$\beta$ is dominated by the interstellar component. For stars colder than $\sim30\,000$\,K, only Ly$\alpha$ can be used for a reliable interstellar H\1\ column density determination, as the stellar absorption starts to dominate the line profile.}
\label{fig:hi}       
\end{figure*}

On the other hand, since low-order Lyman lines (Ly$\alpha$, Ly$\beta$) are characterized by strong interstellar damping wings, interstellar H\1\ dominates the global shape. For instance, Ly$\beta$ can be used to constrain the interstellar H\1\ column density if the stellar temperature is $>30\,000$\,K (Fig.\,\ref{fig:hi}). 

As a summary, the profile of Ly$\alpha$ is a good constrain on $N$(H\1) regardless of the burst age while the profile of Ly$\beta$ can be used only for a dominant stellar population warmer than 30\,000\,K. Higher-order Lyman line profiles provide only little constraints to $N$(H\1) if the signal-to-noise ratio is less than about $\sim50$.

\section{Results}
\label{sec:results}

\begin{figure}
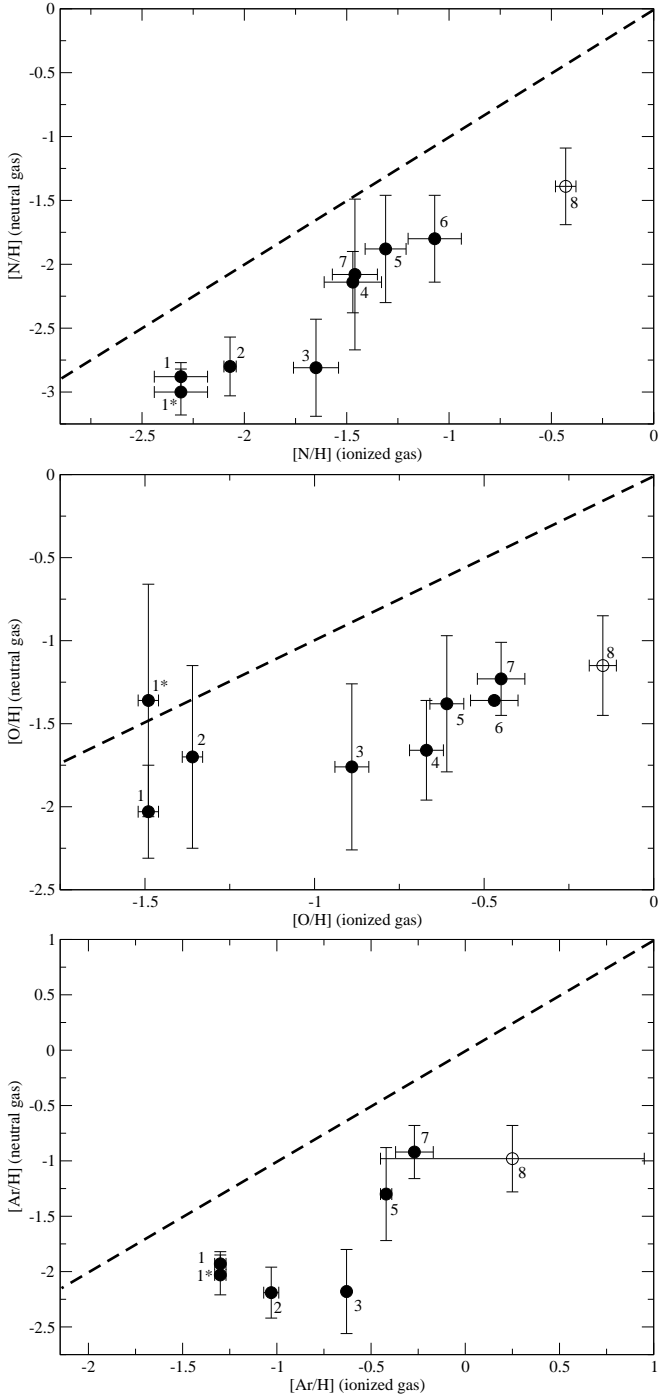

\resizebox{0.7\columnwidth}{!}{  \includegraphics*{fig4a.eps} }
\resizebox{0.7\columnwidth}{!}{  \includegraphics*{fig4b.eps} }
\resizebox{0.7\columnwidth}{!}{  \includegraphics*{fig4c.eps} }
\caption{Abundances of N, O, and Ar in the neutral gas and in the ionized gas (figure taken from Lebouteiller {\em et al.\ }2009). The dashed line indicates the 1:1 ratio. Labels: (1*) and (1) I\,Zw\,18 (see text for details), (2) SBS\,0335--052, (3) I\,Zw\,36, (4) Mark\,59, (5) Pox\,36, (6) NGC\,625, (7) NGC\,1705, (8) NGC\,604/M\,33. }
\label{fig:compa1}       
\end{figure}

Results are shown in Fig.\,\ref{fig:compa1}, where the abundances of N, O, Ar, and Fe in the neutral gas of BCDs are compared to the abundances in the ionized gas of the H\2\ regions. The same dataset was used by Aloisi {\em et al.\ } (2003) and Lecavelier {\em et al.\ }(2004) to investigate I\,Zw\,18 (data points (1) and (1*) respectively). Lecavelier {\em et al.\ } find O\1/H\1$=-4.7^{+0.8}_{-0.6}$ which is consistent with the O/H ratio observed $-0.6$ in the H\2\ regions (all uncertainties are 2-$\sigma$) while Aloisi {\em et al.\ }report a significantly different value with O\1/H\1$=-5.4\pm0.3$, a discrepancy with the former result that awaits for clarification. Hereafter, we define the underabundance of an element X in the neutral has as:
\begin{equation}
\delta_{\rm H\,I}({\rm X}) = {\rm [X/H]}_{\rm H\,II} - {\rm [X/H]}_{\rm H\,I},
\end{equation}
where [X/H]$_{\rm H\,II}$ is the abundance in the ionized gas of the H\2\ regions and [X/H]$_{\rm H\,I}$ is the abundance in the neutral gas. 

Examination of [Ar/H] and [N/H] in Fig.\,\ref{fig:compa1} reveals the possible existence of a plateau for the lowest-metallicity BCDs. The data on [O/H] are also consistent with the existence of a metallicity plateau located at $12+\log {\rm (O/H)} \sim 7.0$ ([O/H] $= -1.7$), although the error bars are considerably larger. Part of the dispersion of [O/H] is probably due to saturation effects which could not be avoided in some objects. Some of the dispersion of [N/H] and [Ar/H] is probably due to ionization corrections. 

Two fundamental results can be drawn from Fig.\,\ref{fig:compa1}: (1) the abundances in the neutral gas show that this phase has been already enriched in heavy elements, and (2) the abundances in the neutral gas are systematically equal or lower than those in the ionized gas ($\delta_{\rm H\,I}({\rm X})>0$). These results suggest that the neutral envelope of BCDs is participating in the dispersion and mixing of heavy elements produced in massive stars. However, the neutral phase seems to be less chemically evolved than the ionized gas associated with the current star-formation episode.

The interpretation of abundances in BCDs has been limited so far by the lack of knowledge about the origin of the neutral gas whose absorption is detected in the FUV. While absorption-lines can arise in the cold and warm neutral medium, it must be kept in mind that there is a bias in the FUV to observe dust-free lines of sight toward the most massive stars. The direct comparison with the H\1\ gas detected \textit{via} the 21\,cm emission in the BCDs is also hampered by the fact that FUV observations only probe gas in front of stellar clusters while radio observations (in emission) probe the whole system. We discuss in the following the properties of the FUV lines of sight by the means of depletion calculation.

\section{Depletion on dust grains}
\label{sec:depletion}

Depletion concerns mostly Si, Fe, and to a minor extent, P. The depletion strength in the neutral envelope of BCDs is largely unknown, mostly because of the lack of information on the FUV lines of sight and because of the low metal content of the gas. This is an important parameter to account for to estimate the fraction of the metal underabundance in the neutral gas genuinely due to the chemical evolution.

The observed abundance of an element X can be expressed as:
\begin{equation}
{\rm [X/H]} \approx \log(Z) + {\rm [X/H]}_0 + A_{\rm X} F_*,
\end{equation}
where [X/H] is the abundance respect to the solar value, i.e., $\log({\rm X/H})-\log({\rm X/H})_\odot$. $Z$ is the intrinsic metallicity of the gas in solar units, [X/H]$_0$ corresponds to the initial (minimal) depletion, $A_{\rm X}$ ($<0$) is the propensity of a given element to be depleted and $F_*$ is the sightline collective depletion strength, following the definitions of Jenkins (2009). It must be noted that [X/H]$_0$ depends on metallicity, being likely closer to $0$ in metal-poor environments because of the lower dust-to-gas ratio (Hirashita {\em et al.\ }2002; Lisenfeld \& Ferrara 1998). 

\begin{figure}
\rotatebox{-90}{\resizebox{0.75\columnwidth}{!}{  \includegraphics{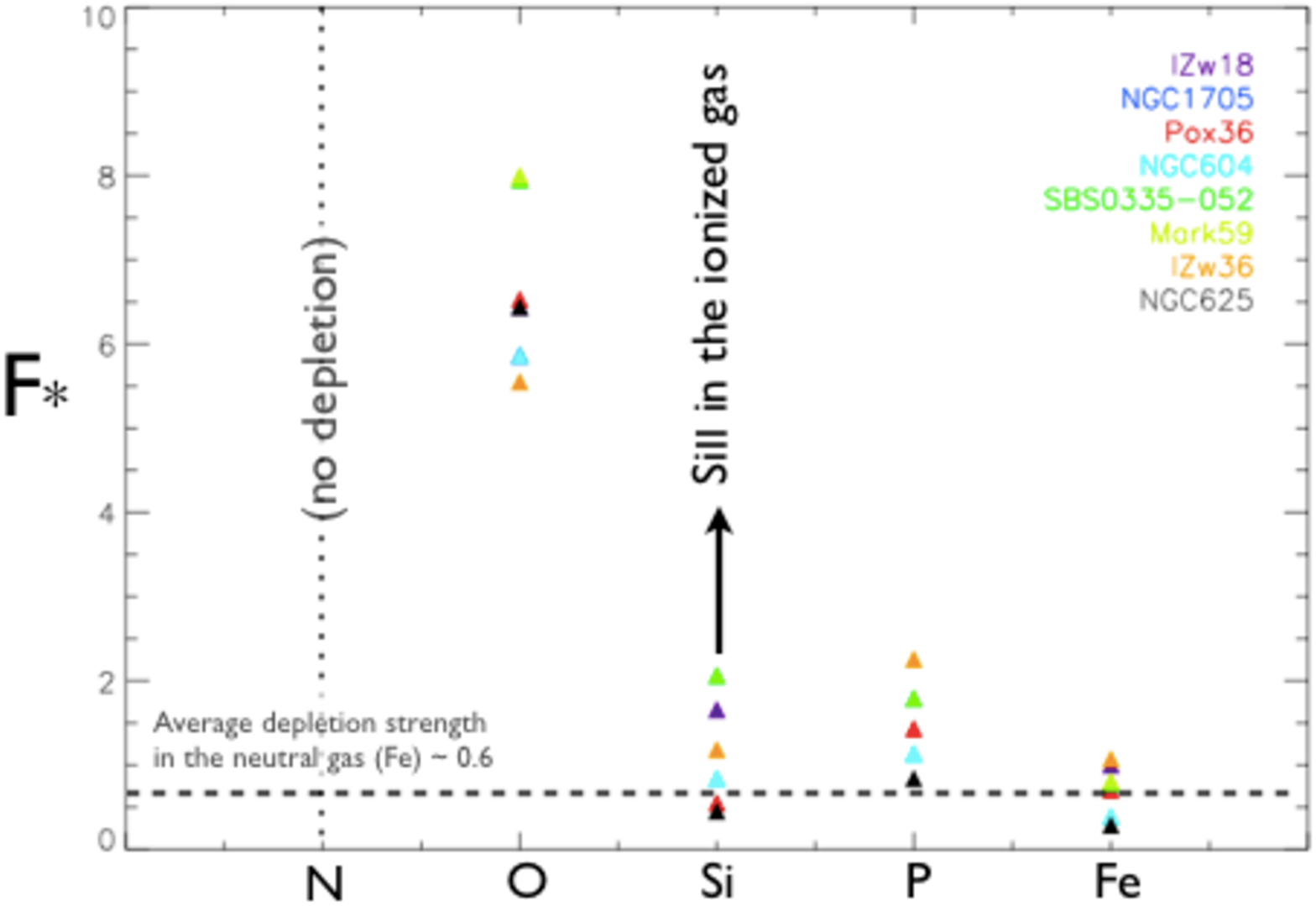} } }
\caption{Depletion strength in the neutral gas assuming that it is the parameter dominating the observed abundances in the neutral gas. This is an upper limit since we assumed a solar metallicity and the neutral gas is intrinsically metal-poor (see text).}
\label{fig:depletion}       
\end{figure}

For simplicity, we make the assumption that the metallicity of the neutral gas is solar $\log(Z)=0$ and try to find the depletion strength $F_*$ agreeing with the observations. This is equivalent to assuming that $\delta_{\rm H\,I}$ is dominated by depletion on dust grains rather than a different chemical enrichment. Figure\,\ref{fig:depletion} shows that $F_*$ in BCDs would then be  comprised between 0.2 and 1.1 (based on the use of the most refractory element, Fe). This is a strong upper limit because we assumed a solar metallicity. In practice, to recover the observed metallicity in the ionized gas, $F_*$ would be essentially null. The fact that $F_*=0$ would imply that the lines of sight are essentially dust free, intersecting UV transparent diffuse clouds. This is of course compatible with the selection effect of lines of sight in the sense that the most opaque clouds are invisible in the UV. 

If $F_*$ is indeed null, underabundances in the neutral gas are not due to depletion on dust grains, as it is already suggested in Fig.\,\ref{fig:compa1} by the study of N and Ar (only little depleted). The observed abundances of N, O, and Ar can only be explained by a genuine chemical underabundance in the neutral gas. 

\begin{figure}
\resizebox{0.95\columnwidth}{!}{  \includegraphics*{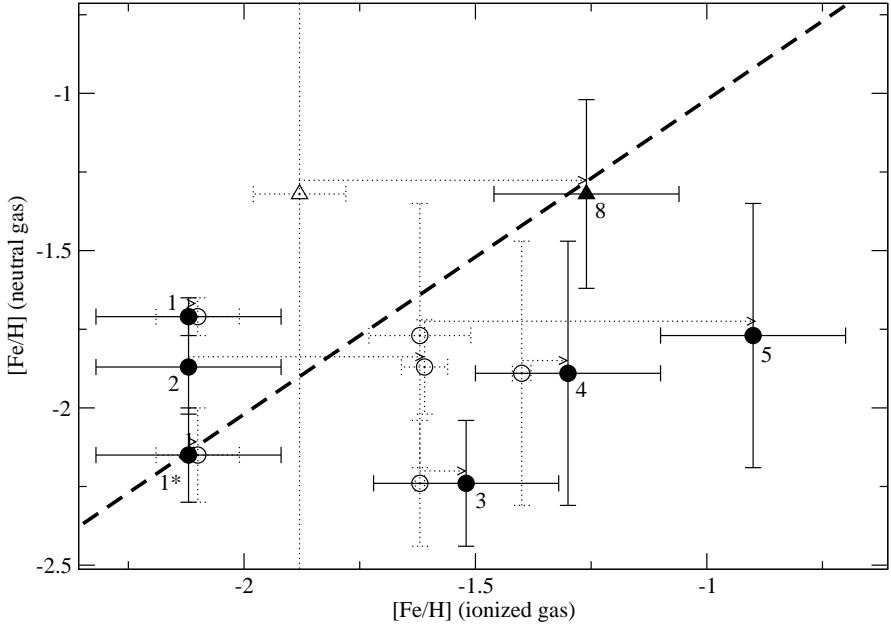} }
\caption{Abundances of Fe in the neutral and ionized gas. See Fig\,\ref{fig:compa1} the plot description. Grey points correspond to the [Fe/H] in the ionized gas (observed values) while black dots include a correction to account for depletion in the ionized gas (figure taken from Lebouteiller {\em et al.\ }2009).  }
\label{fig:compa2}       
\end{figure}

The situation for Fe requires further interpretation. Fe is known to deplete significantly in the \textit{ionized gas}, even in metal-poor objects. In the Milky Way, it is thought that about 2/3 of the iron in the ISM was produced by SNe type Ia involving low-mass stars. Rodriguez \& Esteban (2004) and Izotov {\em et al.\ }(2006) observed that the [O/Fe] ratio increases with metallicity in emission-line galaxies, being consistent with depletion of Fe on dust grains in the ionized gas, and with iron production in SN type II in those galaxies. Hence we could be in a situation where Fe is less depleted in the neutral gas $-$ probed along dust-free lines of sight $-$ than in the ionized gas. Correcting for depletion of Fe in the ionized gas leads to similar abundance patterns to the other elements, i.e., an underabundance in the neutral gas (see Fig.\,\ref{fig:compa2} and Lebouteiller {\em et al.\ }2009).

\section{Discussion}
\label{sec:discussion}

We now assume that all the problems related to the chemical abundance determinations in the neutral gas are solved. These include ionization correction, line saturation effects, depletion in dust grains, contamination by stellar photospheres (for H\1). Can the results be understood in a chemical evolution scenario involving both the ionized gas and the neutral gas?

\subsection{Abundance discontinuity}
\label{sec:discon}

Considering the mass of metals released by a starburst, several hypotheses come into play, as explained in Lebouteiller {\em et al.\ }(2008):
\begin{itemize}
\item All the metals mix in the H\2\ region alone, leading to unrealistically high abundances (see introduction).
\item All the metals mix in the H\1\ envelope. Given the large mass of atomic hydrogen, the metallicity of the neutral gas would increase by $\sim5$\% (calculation made for Pox\,36). 
\item Only a fraction of the metals mixes within the H\2\ gas, and the remaining mix within the H\1\ gas. It is possible that the H\1\ close to the clusters is enriched first. Lebouteiller {\em et al.\ }(2009) calculated that if metals mix with 10\% of the H\1\ in Pox\,36, the enriched H\1\ would acquire the metallicity observed in the ionized gas of the H\2\ regions.
\end{itemize}

The amount of metals released by a single starburst episode cannot be mixed entirely within the H\2\ gas since it would lead to unrealistic abundances. The minimal enrichment therefore applies to only a fraction of the expelled metals. A possible scenario might thus include a local enrichment in the H\2\ region of a $-$ small $-$ fraction of the metals released by the current star-formation episode while most of the metals mix in the H\1\ envelope. In order to reproduce the observed abundance discontinuity between the 2 phases, the fraction of metals mixing in the neutral gas has to satisfy to the condition:
\begin{equation}
f_n = \frac{1}{1 + 10^{\delta_{\rm H\,I}} \frac{M({\rm H\,II})}{M({\rm H\,I})} },
\end{equation}
where $M$(H\2) and $M$(H\1) are the masses of the ionized gas and neutral gas respectively (see Lebouteiller {\em et al.\ }2009) and where $\delta_{\rm H\,I}$ is the abundance discontinuity between the ionized gas and neutral gas as defined in Sect.\,\ref{sec:results}. 

In Pox\,36, $\delta_{\rm H\,I}\approx10$ translates into $f_n$ of about 1\%. In I\,Zw\,18, $\delta_{\rm H\,I}\approx0$ (following the result of Lecavelier {\em et al.\ }2004) translates into a fraction $f_n$ of 5\%. The large difference in behavior between $f_n$ and  $\delta_{\rm H\,I}$ in these 2 galaxies is due to the fact that the ratio of ionized gas mass over neutral gas mass is much larger in I\,Zw\,18 than in Pox\,36.

\subsection{Metallicity threshold?}

The neutral gas of BCDs has already been enriched with metals up to an amount of $>1/50$\,Z$_\odot$. Could this value be a metallicity threshold? The hypothesis of a minimal metallicity has been already suggested by Kunth \& Sargent (1986) to explain the difficulty to find extremely-metal poor star-forming objects. In these objects, the current star-formation episode could pollute the ISM by heavy elements released by stellar winds and supernov\ae.  It is interesting to notice that the apparent threshold of $>1/50$\,Z$_\odot$ in the neutral gas of BCDs is also the metallicity of the most metal-poor galaxy known, I\,Zw\,18. It is conceivable that the results are related. However, they imply considerably different (both spatial and time) scales for the dispersion and for the mixing.

Interestingly, Telfer {\em et al.\ }(2002) found that the intergalactic medium (IGM) in the redshift range $1.6<z<2.9$ has a metallicity $12+\log ({\rm O/H})$ between $6.7$ and $7.6$, which includes the floor metallicity observed in BCDs. Hence it appears that the neutral and ionized ISM of the lowest metallicity galaxies could well be characterized by a default enrichment rather than by previous star-formation episodes. In that view, BCDs could have started to form stars only "recently" (around $\sim8-9$\,Gyrs ago), from an H\1\ cloud already at the IGM metallicity. This has important consequences on the number of starburst episodes that occurred until now. For example, assuming $f_n=5$\%\ in I\,Zw\,18 (to reproduce the result $\delta_{\rm H\,I}=0$, Sect.\,\ref{sec:discon}), 3 bursts are required to explain the observed metallicity if starting from a null metallicity (calculation made using yields from Meynet \& Maeder (2002) and assuming interphase mixing between 2 consecutive bursts). However, a single burst is required if starting from $Z<1/50$\sol. Further observations and analysis of BCDs with metallicities lower than $\sim1/8$\,Z\,$_\odot$ will be needed to confirm the existence of this plateau.

\subsection{Metals from star-forming dwarfs: retention or ejection?}

Since a local enrichment of 100\%\ of the products released by a starburst would lead to unrealistically high abundances, 2 possibilities remain: (1) metals disperse and mix in another phase (Sect.\,\ref{sec:discon}), or (2) most of the metals are ejected into the IGM. We now discuss the parameters playing a role in the ejection/retention of metals.

As they end their lives massive stars explode as a supernov\ae. The energy output from a SN is over a short period, comparable to that of a whole galaxy. In a galaxy with a high local star formation rate, the collective action of supernov\ae\ may lead to a galactic superwind, which may cause loss of gas. Stellar winds can also contribute to the energetics of the ISM at the very early stage of a starburst (Leitherer {\em et al.\ }1992). The relative importance of stellar winds compared to SNe increases with metallicity. A continuous wind proportional to the star formation rate has been applied in models predicting the evolution of starburst galaxies. But since different elements are produced on different timescales, it has been proposed that only certain elements are lost (or in different proportions) hence reducing the effective net yield of those metals as compared to a simple chemical evolution model (Matteucci \& Chiosi 1983; Edmunds 1990). The SNe involved in such a wind are likely to be of type II because type Ia SNe explode in isolation and will less likely trigger chimneys from which metals can be ejected out of the plane of a galaxy. In this framework O and part of Fe are lost while He and N (largely produced by intermediate stars) are not. This would result in a cosmic dispersion in element ratios such as N/O between galaxies that have experienced mass loss and those that have not. In a dwarf galaxy which has a weaker gravitational potential, these effects may result in gas loss from the galaxy unless as we argue below the presence of a low H\1\ density halo acts as a barrier.

Galactic winds have been observationally investigated in dwarf galaxies (e.g., Marlowe {\em et al.\ }1995; Martin(1996; Martin(1998) and more recently with the advent of the Chandra satellite. Lequeux {\em et al.\ }(1995), Kunth {\em et al.\ } (1998), and Mas-Hesse {\em et al.\ }(2003) have shown that the escape of the Ly$\alpha$ photons in star-forming galaxies strongly depends on the dynamical properties of their interstellar medium. The Lyman alpha profile in the BCG Haro 2 indicates a superwind of at least 200\,km\,s$^{-1}$, carrying a mass of  $\approx10^7$\,M\sol, which can be independently traced from the H$\alpha$ component (Legrand {\em et al.\ }1997). However, high speed winds do not necessary carry a lot of mass. Martin (1996) argues that a bubble seen in IZw 18 (see also Petrosian {\em et al.\ }1997) will ultimately blow-out together with its hot gas component. Although little is known about the interactions between the evolving supernova remnants, massive stellar bubbles and the ISM it is possible that an outflow takes the fresh metals with it and in some cases leaves a galaxy totally cleaned of gas. This scenario clearly contradicts the self enrichment picture $-$ unless a fraction of hot gas rapidly cools down $-$. But since we see some hot gas outside of the H\2\ regions the question remains whether this gas will leave a galaxy or simply stay around in the halo?

Model calculations developed by Silich \& Tenorio-Tagle (1998) predict that superbubbles in amorphous dwarf galaxies must have already undergone blowout and are presently evolving into an extended low-density halo. This should inhibit the loss of the swept-up and processed matter into the IGM. Recent Chandra X-rays observations of young starbursts indicate some possible metal losses from disks (Martin {\em et al.\ }2002) but not in the case of NGC\,4449 with an extended H\1\ halo of around $40$\,kpc (Summers {\em et al.\ }2003). In a starburst galaxy, newly processed elements are produced within a region of $\sim100$\,pc size. During the supernova phase the continuous energy input rate from coeval starbursts or continuous star-forming episodes, maintains the temperature of the ejected matter above the recombination limit ($T\sim10^6$\,K) allowing superbubbles to reach dimensions in excess of $1$\,kpc.

\begin{figure}[t!]
\rotatebox{-90}{\resizebox{0.8\columnwidth}{!}{  \includegraphics*{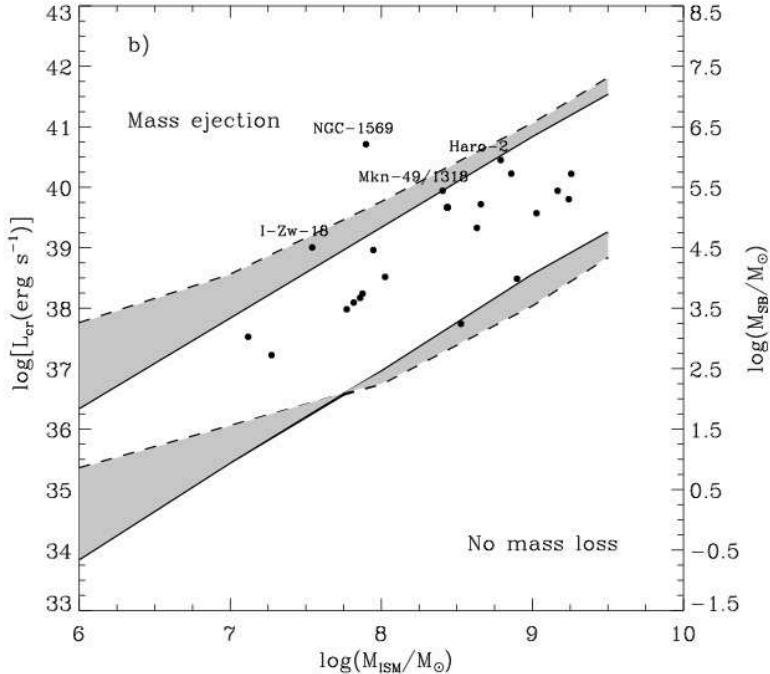} } }
\caption{Log of the critical mechanical luminosity (left-side axis) and mass of the star cluster $M_{\rm SB}$ (right-side axis), required to eject matter from galaxies as a function of $M_{\rm ISM}$ (figure taken from Legrand {\em et al.\ }2001). Lower limit estimates are shown for galaxies with extreme ISM density distributions: flattened disks (lower two lines), and spherical galaxies without rotation (upper lines), for two values of the intergalactic pressure $P_{\rm IGM}/k = 1$\,cm$^{-3}$\,K (solid lines) and $P_{\rm IGM}/k = 100$\,cm$^{-3}$\,K (dashed lines).}
\label{fig:legrand}       
\end{figure}

The mechanical energy released during a starburst episode accelerates the interstellar medium gas and generates gas flows. The properties and evolution of these flows ultimately determine the fate of the newly formed metals and the manner they mix with the original interstellar medium. The presence of outflows may indicate, at first, that supernova products and even the whole of the interstellar medium may be easily ejected from the host dwarf systems, causing the contamination of the intra-cluster medium (Dekel \& Silk 1986; De Young \& Heckman1994). This type of assumption is currently blindly used by cosmologists in their model calculations. However the indisputable presence of metals in galaxies implies that supernova products are not completely lost in all cases (Silich \& Tenorio-Tagle 2001). Legrand {\em et al.\ }(2001) have compared Silich \& Tenorio-Tagle (2001) theoretical estimates with some well-studied starburst galaxies. They have worked out three different possible star formation history scenarios that assume either a very young coeval starburst or extended phases of star formation of 40\,Myr or 14\,Gyr and inferred the expected energy input rate for each galaxy using the H$\alpha$ luminosity and/or the observed metallicity. Values of the derived mechanical energy injection rate in the three considered cases were compared with the hydrodynamical models predictions of Mac Low \& Ferrara (1999) and Silich \& Tenorio-Tagle (2001). Detailed calculations and accompanying figures are given in Legrand {\em et al.\ }(2001); we only present in Fig.\,\ref{fig:legrand} the resulting plot for the extended phases of star formation of 40\,Myr that might be the most realistic case for starburst galaxies.

The net result (see Fig.\,\ref{fig:legrand}) is that all galaxies lie above the lower limit first derived by Mac Low \& Ferrara (1999) for the ejection of metals out of flattened disk-like ISM density distributions energized while most are \textit{below} the limit for the low density halo picture. Thus the mass of the extended low density halo efficiently acts as the barrier against metal ejection into the IGM.

Disk-like models clearly require less energy to eject their metals into the intergalactic medium because the amount of blown out interstellar gas that can open a channel into the intergalactic medium is much smaller than in the spherically symmetric limit, where all of the metal-enriched ISM has to be accelerated to reach the galaxy boundary. Predicted haloes, despite acting as a barrier to the loss of the new metals, have rather low densities ($<n_{\rm halo} \sim10^{-3}$\,cm$^-3$) and thus have a long recombination time ($t_{\rm rec} = 1/(\alpha n_{\rm halo})$; where $\alpha$ is the recombination coefficient) that can easily exceed the lifetime of the H\2\ region ($t_{\rm HII} = 10^7$\,yr) developed by the starburst. In such a case, these haloes may remain undetected at radio and optical frequencies until large volumes are collected into the expanding supershells.

\subsection{Are BCDs special?}

Bowen {\em et al.\ }(2005) found similar abundances in the neutral and ionized gas of the dwarf spiral galaxy SBS\,1543+593. This results poses a fundamental question: is the metal deficiency of the H\1\ gas specific to BCDs? Rotation in spirals provide an efficient mixing throughout the galaxy. Furthermore, star-formation keeps occurring along the spiral arms. In opposition, BCDs have weak velocity field, and star-formation usually occurs in one or several knots (e.g., Papaderos {\em et al.\ }1996). The different mode of star-formation could be responsible for substantial differences in the dispersal and mixing mechanisms. Lebouteiller {\em et al.\ }(2006) found that the neutral phase toward NGC\,604 in the spiral galaxy M\,33  is more metal-poor than the ionized gas of the H\2\ region, i.e., being consistent with results in BCDs, but the contamination by high-velocity clouds along the lines of sight similar to the ones observed around the Milky Way complicates the interpretation. It seems necessary to study the metal abundance in the neutral gas of dwarf irregular galaxies that are not BCDs. Studies of other star-forming objects ought to be performed, which is now possible thanks to the interactive database tool designed by Desert {\em et al.\ }(2006a) and Desert {\em et al.\ }(2006b). 

\subsection{Perspectives}

One of the most important complication in the studies of absorption-line spectra of galaxies is the presence of unseen components, possibly saturated. This caveat is usually solved by the use of the weakest observed lines. In that vein, the HST/COS instrument enables the sulfur abundance determination in the neutral phase through the interestingly weak S\2\ $\lambda1256$ multiplet (Pettini \& Lipman 1995). The oxygen abundance will be measured with a better accuracy than with \textit{FUSE} thanks to the O\1\ $\lambda1356$ line (Meyer {\em et al.\ }1998). Furthermore, the examination of phosphorus lines will confirm the possible use of phosphorus abundance as a metallicity tracer as suggested by Lebouteiller {\em et al.\ }(2005). This is particularly important for the analysis of the \textit{FUSE} database since O\1\ lines are generally saturated or close to saturation. Finally, it must be stressed that the ideal case study for deriving abundances in the neutral gas of BCDs is to observe the ISM toward a quasar line of sight (Kunth \& Sargent 1986). The method was successfully used in the dwarf spiral galaxy SBS\,1543+593 by Bowen {\em et al.\ }(2005) but unfortunately, no alignment BCD-quasar has been observed so far. The study of such an alignment would cancel the selection effect in \textit{FUSE} spectra to probe gas toward star-forming regions within the galaxy. 

As far as chemical models are concerned, Recchi {\em et al.\ }(2004) examined the case of I\,Zw\,18 for which they found that abundances in the H\1\ medium (defined by a temperature lower than $7\,000$\,K) are similar to those in the H\2\ gas. Hence, although some physics is still missing in the models (radiative transfer, clumpy ISM, dust phase), it seems that FUSE results still cannot be reproduced. 

\section{Conclusion}

The lower metallicity observed in the H\1\ gas of BCDs as compared to the ionized gas of their H\2\ regions remains a puzzle. We have discussed many effects that may affect the determination of the neutral gas abundances such as ionization corrections, depletion on dust grains, stellar contamination, and line saturation. The underabundance of a factor 10 found in the H\1\ region of several BCDs is however so dramatic that it seems difficult all these effects can make us miss the detection of 90\%\ of the metals. 

It is possible to reproduce the abundance discontinuity in the 2 phases assuming that only a fraction (typically a few percents) of the metals released by the current starburst episode mix within the ionized gas, while most of them mix with the neutral envelope. The considerable mass of H\1\ results in only a slight metallicity increase, even with most of the metals mixing within. The main parameter therefore concerns the ratio of ionized-to-neutral gas, in the sense that a galaxy with a large fraction of its total gas mass ionized is more likely to show similar abundances in the 2 phases. Note that his hypothesis is simple speculation at this stage and it has to be tested against dispersion spatial scales and against mixing timescales.

It is essential to confirm the existence of a metallicity floor. The lowest metallicities observed in the neutral gas of BCDs agree with the metallicity in I\,Zw\,18 $-$ the most metal-poor star-forming galaxy known $-$, suggesting a minimal enrichment regardless of the ISM phase. Such a minimal enrichment seems to be in agreement with the IGM metallicity around a redshift $\sim2$, so that BCDs might have formed only "recently" ($<10$\,Gyrs ago). 

%


%

%

\end{document}